\documentclass[fleqn,usenatbib]{mnras}

\usepackage[T1]{fontenc}

\DeclareRobustCommand{\VAN}[3]{#2}
\let\VANthebibliography\thebibliography
\def\thebibliography{\DeclareRobustCommand{\VAN}[3]{##3}\VANthebibliography}

\usepackage{graphicx}	
\usepackage{amsmath}	
\usepackage{amssymb}	
\usepackage{xcolor}

\newcommand{\Msunpc}{\,M$_\odot$\,pc$^{-1}$} 
\newcommand{\kms}{\,km\,s$^{-1}$} 
\newcommand{\K}{\,K} 
\newcommand{\Myr}{\,Myr} 
\newcommand{\pc}{\,pc} 

\allowdisplaybreaks

\title[Suppressing the end dominated collapse]{Filament fragmentation: Density gradients suppress end dominated collapse}

\author[E. Hoemann et al.]{
Elena Hoemann,$^{1,2}$\thanks{E-mail: hoemann@usm.lmu.de}
Stefan Heigl$^{1,3}$
and Andreas Burkert$^{1,2,3}$
\\
$^{1}$Universitäts-Sternwarte, Ludwig-Maximilians-Universität München, Scheinerstr. 1, 81679 Munich, Germany \\
$^{2}$Max-Planck Institute for Extraterrestrial Physics, Giessenbacherstr. 1, 85748 Garching, Germany \\
$^{3}$Excellence Cluster ORIGINS, Boltzmannstrasse 2, 85748 Garching, Germany
}

\date{Accepted 2023 August 17. Received 2023 August 16; in original form 2023 July 20
}

\pubyear{2023}

\usepackage{newtxtext,newtxmath}
\begin{document}
\label{firstpage}
\pagerange{\pageref{firstpage}--\pageref{lastpage}}
\maketitle

\begin{abstract}
The onset of star formation is set by the collapse of filaments in the interstellar medium. From a theoretical point of view, an isolated cylindrical filament forms cores via the edge effect. Due to the self-gravity of a filament, the strong increase in acceleration at both ends leads to a pile-up of matter which collapses into cores. However, this effect is rarely observed. Most theoretical models consider a sharp density cut-off at the edge of the filament, whereas a smoother transition is more realistic and would also decrease the acceleration at the ends of the filament. We show that the edge effect can be significantly slowed down by a density gradient,  although not completely avoided. However, this allows perturbations inside the filament to grow faster than the edge. We determine the critical density gradient for which the timescales are equal and find it to be of the order of several times the filament radius. Hence, the density gradient at the ends of a filament is an essential parameter for fragmentation and the low rate of observed cases of the edge effect could be naturally explained by shallow gradients.
\end{abstract}

\begin{keywords}
stars:formation -- ISM:kinematics and dynamics -- ISM:structure
\end{keywords}



\section{Introduction}

  Observations show that the molecular interstellar medium is pervaded by filaments \citep{Schneider1979, Andre2010, Arzoumanian2013, Andre2014}. They are key in understanding the formation of stars since prestellar cores are mainly found within these cold, dense, filamentary structures \citep{Arzoumanian2011, Koenyves2015}. They span several orders of magnitude \citep{Hacar2022} from hundreds of parsecs in length down to the sub-parsec regime \citep{Molinari2010, Andre2010, Hacar2013, Goodman2014, Mattern2018, Schmiedeke2021}. Besides their obvious connection to star formation, there still remain many open questions on the formation, evolution and collapse of filaments.

  Particularly the formation of cores and their consequent collapse is interesting since this is the stage of early star formation. For isolated filaments, fragmentation is supposed to happen via two competing processes: the edge effect \citep{Bastien1983, Burkert2004, Pon_2012} and the growth of perturbations \citep{Stodolkiewicz1963, Nagasawa1987, Fischera2012}. The edge effect consists of creating cores at the ends of the filament during the overall filament collapse. Since there is no density distribution for which the filament is in hydrostatic equilibrium along its main axis, it will collapse longitudinally due to self-gravity. This gravitational acceleration has a sharp increase in the end regions of the filament, and thus matter is swept up at the ends which forms a core and collapses after accumulating enough mass \citep{Burkert2004, Hartmann2007, Li2016}. Contrary to this theoretical expectation, such pronounced ends are rarely observed \citep{Zernickel2013, Kainulainen2016, Dewangan_2019, Bhadari_2020, Yuan2020, Cheng2021}, which leads to the question of why we do not observe end-dominated filaments more often.

  For magnetised, disc-like clouds the formation of an outer ring is the pendant to the edge effect. \citet{Li2001} already found that the creation of a ring is correlated to the profile of the disc and the sound speed. However, for filaments \citet{Seifried2015} showed that an initial density peak in the centre leads to a centrally-dominated collapse instead of an end-dominated one. In case of short filaments with aspect ratios of 3:1 the longitudinal collapse was also investigated by \cite{Keto2014}. They showed that for low line-masses the filament collapses longitudinally into a central core that begins to go through a complex pattern of oscillations. Although such a longitudinal collapse has not yet been confirmed by observations \citep{Roy2015} the predicted oscillations of the resulting cores are \citep{Redman2006,Aguti2007}. In addition, \cite{Heigl2022} showed that a filament created by a constant inflow region can suppress the edge effect. The constant inflow region rebuilds the filament during the overall filament collapse leading to a density gradient in the end region. This gradient leads to a decrease in the acceleration at the end of the filament which slows down the creation of the edge effect. Although not all filaments are expected to have such high and constant accretion, a smooth transition from the filament end into the surrounding medium is expected. This raises the question of whether a gradient at the end of the filament can stop the edge effect and if so, under which condition. 
  
  In general, cores are much more often detected inside filaments. These can form due to the growth of density perturbations. Although the expected spacing of the cores due to perturbation theory \citep{Nagasawa1987,Larson1985,Inutsuka1992, Gehman1996, Hosseinirad2017} is mostly not observed, several filaments with regularly spaced cores were detected \citep{Tafalla2015,Zhang2020}.
  
  In this paper, we show that the edge effect itself cannot be stopped by any density profile at the end region of the filament. However, the collapse can be slowed down by a density gradient, such that perturbations can grow faster inside the filament. This is a natural explanation why filaments rarely show an edge effect. We present an analytic model for the conditions under which the edge should be dominant and under which conditions perturbations can grow faster and confirm it by numerical simulations.

  The paper is organized as follows: First, the basic principle of hydrostatic filaments, the edge effect and perturbations are discussed in Section \ref{sec:basic_principles}. Then we argue why the edge effect can never be avoided completely in Section \ref{sec:edge_supression}. The derivation of the critical density gradient for which perturbations within the filament grow faster than at the edge is presented in Section \ref{sec:critical_denisty_gradient}, followed by the validation by simulations in Section \ref{sec:simulation}. Finally, the results are discussed and conclusions are drawn in Sections \ref{sec:discussion} \& \ref{sec:conclusion}.

\section{Basic principles}
\label{sec:basic_principles}

  \subsection{Hydrostatic filaments}
  
  Considering filaments as isothermal cylinders, a hydrostatic solution for the radial profile was already found by \citet{Stodolkiewicz1963} and \citet{Ostriker1964}:
  \begin{align}
      \rho (r) = \rho _\mathrm{c} \left[ 1 + \left( \frac{r}{H} \right)^2 \right]^{-2}
  \end{align}
  with $H$ the scale height:
  \begin{align}
      H^2 = \frac{2c_\mathrm{s}^2}{\uppi G \rho_\mathrm{c}}
  \end{align}
  and $c_\mathrm{s}$ being the sound speed (0.19\kms for a mean molecular weight of 2.36 and a temperature of 10\,K), $G$ the gravitational constant and $\rho_\mathrm{c}$ the central density of the filament. The filament is radially constrained by the pressure equilibrium between the boundary pressure and the external pressure $P_{\mathrm{ext}} = P_{\mathrm{b}}$. Its line-mass is given by its overall mass divided by its length:
  \begin{align}
      \mu =  \frac{M}{l}.
  \end{align}
  Integrating the density distribution radially until infinity results in the maximum line-mass for which a hydrostatic solution exists, given by
  \begin{align}
      \mu _{\mathrm{crit}} = \frac{2 c_\mathrm{s}^2}{G} \approx 16.4 \text{\Msunpc},
  \end{align}
  above which all filaments would start to collapse radially. The value is calculated for 10\K. The criticality $f$ is then the ratio of the actual line-mass divided by the critical line-mass
  \begin{align}
      f = \frac{\mu}{\mu_{\mathrm{crit}}}.
  \end{align}
  The radial boundary density of the filament can be calculated by
  \begin{align}
      \rho _\mathrm{b} = \rho _\mathrm{c} (1-f)^2
  \end{align}
  and the radius of this boundary is given by
  \begin{align}
      R = H \left( \frac{f}{1-f} \right) ^{1/2}.
  \end{align}
  
  \subsection{Edge effect}

  Since there is no hydrostatic solution for the density distribution along the main axis of a filament, the filament is expected to collapse longitudinally. The acceleration along a filament due to its self-gravity was already investigated by \citet{Burkert2004}:
  \begin{align} \label{eq:accelerationCollapseFilament}
	a = -2 \uppi G \bar{\rho} \left[ 2z - \sqrt{\left( \frac{l}{2} + z \right)^2+R^2} + \sqrt{\left( \frac{l}{2} - z \right)^2+R^2} \right].
  \end{align}
  with $\bar{\rho}$ the mean density and $z$ the position along the filament. The steep increase of $a$ at the end leads to a large velocity inwards along the main axis and as a result to a pileup of matter in these regions. The resulting clumps will begin to collapse onto themselves when they reach the critical line mass. At the same time, the end clumps still move inwards towards the centre of the filament, destroying the filament \citep{Bastien1983,Toala2011,Pon_2012,Clarke2015}. This is the so-called edge effect \citep{Burkert2004}. In an earlier paper \citep{Hoemann2021} we investigated on which timescale the ends of a filament are formed:
  \begin{align} 
      t_{\mathrm{edge}} &= \sqrt{\left( \frac{1}{f}-1\right) \frac{2\kappa R}{|a_{\mathrm{cm}}|}} \nonumber\\ 
      &= \sqrt{ \frac{1.69\times 10^{-20}\mathrm{\,g\,cm^{-3}}}{f\rho _\mathrm{c} }} \mathrm{\Myr}
      \label{eq:edge}
  \end{align}
  with $\kappa=1.66$ and $a_{\mathrm{cm}}$ being the acceleration of the centre of mass of the filaments end region $a_\mathrm{cm}=a( l/2-\kappa R/2 )$. In the beginning, the two cores move in free fall and then are slowed down by ram pressure \citep{Hoemann2022}. With this two-phase approach, the longitudinal collapse timescale of a filament can be determined \citep{Clarke2015, Hoemann2022}
  \begin{align}
      t_{\mathrm{col}} &= \frac{0.42+0.28 A}{\sqrt{G\bar{\rho}}} \\
      &= \frac{0.42+0.28A}{\sqrt{\bar{\rho}/1.50\times10^{-20}\mathrm{\,g\,cm^{-3}}}} \mathrm{\Myr},
  \end{align}
  with $A=L/2R$ being the aspect ratio and $\bar{\rho}$ the mean density of the filament. This is now the overall collapse time, the timescale on which the two end cores move to the centre.
  
  Although the edge effect is predicted by theory, it is only rarely observed \citep{Zernickel2013, Kainulainen2016, Dewangan_2019, Bhadari_2020, Yuan2020, Cheng2021}. This raises the question of why we do not observe the edge effect in most cases. However, most of the former theoretical studies were done with a sharp density cut-off at the ends of the filament which influences the acceleration in the end region. A smoother density gradient could influence and weaken the edge effect significantly allowing perturbations to grow faster. This is what we will investigate in the next sections.
  
  \subsection{Perturbations}
  
  In competition to the edge effect, line-mass perturbations within the filament can also grow. Here we characterize a perturbation of the line-mass by
  \begin{align}
  \label{eq:perturbations}
    f(z) = f_0 \left[1+ \epsilon \cos{\left(\frac{2\uppi z}{\lambda_{\mathrm{dom}}}\right)}\right],
  \end{align}
  with $f_0$ the unperturbed line-mass and z the coordinate along the filament. $z=0$ corresponds to the centre of the perturbation. $\epsilon$ is the perturbation strength which is observed to be typically 0.09 \citep{Roy2015} and $\lambda_{\mathrm{dom}}$ is the dominant fragmentation length \cite[see][Appendix E]{Fischera2012}
  \begin{align}
      \lambda _{\mathrm{dom}} = \left( 6.25 - 6.89 f^2 + 9.18 f^3 - 3.44 f^4 \right) FWHM_{\mathrm{cyl}} \label{eq:lambda}
  \end{align}
  using the full width at half maximum of the cylindrical filament $FWHM_\mathrm{cyl}$. Following a perturbation analysis \citep{Nagasawa1987} the timescale can now be determined on which such a perturbation would grow into a collapsing core \citep[compare][]{Heigl2020,Hoemann2021}
  \begin{align}
  \label{eq:perturbationtimescale}
      t_{\mathrm{pert}} = \tau _{\mathrm{dom}} \log \left[ \left( \frac{1}{f} -1 \right) \frac{1}{\epsilon} \right].
  \end{align}
  Here $\tau_{\mathrm{dom}}$ denotes the dominant growth timescale, which can be determined with \citet{Fischera2012}, Appendix E
  \begin{align}
      \tau _ {\mathrm{dom}} = \left(4.08 - 2.99 f^2 + 1.46 f^3 + 0.40 f^4 \right) / \sqrt{4\pi G \rho _\mathrm{c}}.
  \end{align}
  
  Since most of the cores are found within filaments \citep{Arzoumanian2011, Koenyves2015}, this is expected to be one of the most important channels for core formation, theoretically. 
  
\section{The inevitable edge effect}
\label{sec:edge_supression}

  In former theoretical studies \cite[e.g.][]{Clarke2015, Hoemann2021}, filaments are often considered to have a sharp edge, thus, a hard cut-off at the ends of the filament which makes analytic predictions possible \citep{Burkert2004}. However, a more realistic setup would be a smooth transition to the surrounding gas. Such a density gradient would lead to a decrease in acceleration at the ends of the filament.
  \begin{figure}
      \centering
      \includegraphics[width=\columnwidth]{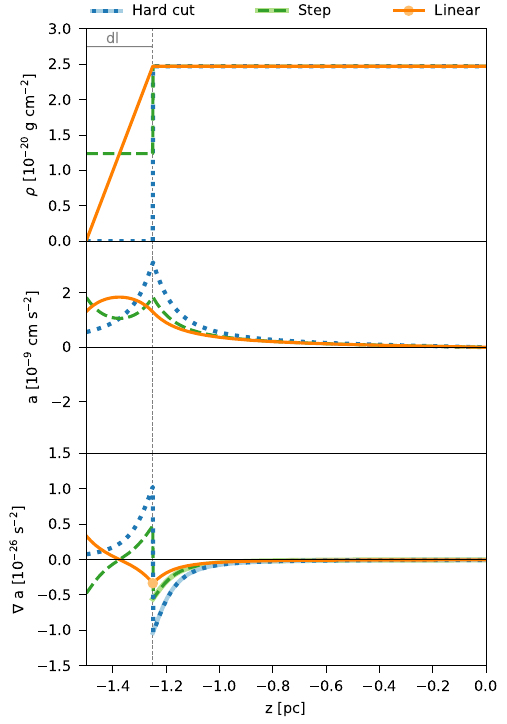}
      \caption{The top row shows the density distribution along the z-axes for three different profiles: A hard cut-off, where the density goes directly to zero after the filament ends, in dotted blue; A step for which the density is cut in half after the end and drops to zero at the edge of the end region, shown in dashed green; A linear end is depicted in orange, where the density transitions more smoothly into the surrounding. In the second row, the corresponding numerically determined acceleration is given and in the last row its gradient. For the gradient, the lighter colours depict the analytic solution which fits the numeric solution well.}
      \label{fig:gradA}
  \end{figure}

  The change in acceleration due to the end profile of the filament is depicted in Figure \ref{fig:gradA}. The top row shows three simple density distributions. The hard cut-off is depicted as a blue dotted line. Here the density drops immediately to the external density at the end of the filament, which is often used for analytic approaches. The green dashed line shows a step edge, where the density halves at the filaments end and only drops to the external density at the edge of the end region. A linear end is given by the solid orange line. In this case, after the end of the filament the density decreases linearly to the external density. We consider the external density to be negligible and set it to zero, to make analytic predictions comparable. The second row depicts the acceleration for each of the profiles and the last row shows the gradient of the acceleration respectively. The acceleration of the hard cut-off can also be determined by Equation \ref{eq:accelerationCollapseFilament}. The calculation was done numerically, but the lines and markers in lighter colours in the last row show the analytical expectation, described in more detail in the following, which fits the numerical solutions. 
  
  Differentiating Equation \ref{eq:accelerationCollapseFilament} leads to the gradient for the hard cut-off, shown in light blue in the lower plot in Figure \ref{fig:gradA} (derivation Appendix \ref{ap:hard_cut}):
  \begin{align}
      \nabla a_{\mathrm{cut}} = 2 \uppi G \rho \left[ \frac{l/2+z}{\sqrt{(l/2+z)^2+R^2}} + \frac{l/2-z}{\sqrt{(l/2-z)^2+R^2}} - 2 \right].
  \end{align}
  For the step function, which is depicted in light green, considering of outside material leads to:
  \begin{align}
      \nabla a_{\mathrm{step}} =& -2 \uppi G \rho \Bigg[ -\frac{1}{2} \frac{l/2+z}{\sqrt{R^2+(l/2+z)^2}} - \frac{1}{2} \frac{l/2-z}{\sqrt{R^2+(l/2-z)^2}} + 2 \nonumber \\
      & - \frac{1}{2} \frac{l/2+dl+z}{\sqrt{R^2+(l/2+dl+z)^2}} - \frac{1}{2} \frac{l/2+dl-z}{\sqrt{R^2+(l/2+dl-z)^2}}\Bigg]
  \end{align}
  $dl$ donates the length of the additional end region, given in Figure \ref{fig:gradA}. The derivation can be found in Appendix \ref{ap:step}. For the linear density distribution at the end, we calculated the gradient of the acceleration only for the filament end since this is the region of interest (derivation in Appendix \ref{ap:linear}):
  \begin{align}
      \nabla a_{\mathrm{lin}} = 2 \uppi G \rho \bigg[ \frac{\sqrt{R^2+dl^2}}{dl} - \frac{R}{dl} + \frac{l}{\sqrt{R^2+l^2}} - 2 \bigg].
       \label{eq:acc_gradient}
  \end{align}
  We checked the limits for consistency. Case 1: $dl \to0$ reproduces the gradient for the hard cut-off:
  \begin{align}
      \lim_{dl\to0} \nabla a_{\mathrm{lin}} = 2 \uppi G \rho \left[ \frac{l}{\sqrt{R^2+l^2}} - 2 \right]
  \end{align}
  and case 2: $dl \to \infty$ has to be zero since it would be an infinite long filament, thus also $l \to \infty$
  \begin{align}
      \lim_{dl,l \to \infty} \nabla a_{\mathrm{lin}} = 0.
  \end{align}
  
  Altogether, the important parameter for the collapse of a filament is the amount of acceleration in the end region, which can be weakened by a smooth density transition. However, every density gradient will at least cause a small peak in the gradient of the acceleration, as can be seen in Equation \ref{eq:acc_gradient}. Thus, during the filament collapse the formation of an edge effect will always be triggered. The amount of acceleration determines how fast the edge grows:
  \begin{align}
      t_{\mathrm{edge}} \propto \frac{1}{\sqrt{a}}.
  \end{align}
  Therefore, the edge effect can be slowed down by a density gradient at the end, reducing the acceleration in this region. However, the collapse timescale of a filament scales with the same dependence on a:
  \begin{align}
      t_{\mathrm{col}} \propto \frac{1}{\sqrt{a}} 
  \end{align}
  which means that if the filament encounters an edge effect with a cut-off, it also encounters an edge effect with any end profile before collapsing, because collapse and edge effect are slowed down by the same amount. Thus, a gradient cannot stop the edge effect, it can just slow it down.
  
  However, there is also a third effect, the growth of perturbations, which plays an important role which we consider in the next section.

\section{Critical density gradient}
\label{sec:critical_denisty_gradient}

  As was shown in the previous section, the edge effect cannot be stopped, but it can be slowed down. The question is, under with circumstances is the slowdown sufficient to make perturbations grow faster than the edge effect. In contrast to the overall collapse timescale, the timescale on which perturbations grow is independent of the acceleration at the ends. Thus, even if the acceleration at the ends is low, perturbations grow nevertheless. Hence, if perturbations grow faster than the end cores, we do not expect to see a dominant edge effect.
  
\begin{figure}
      \centering
      \includegraphics{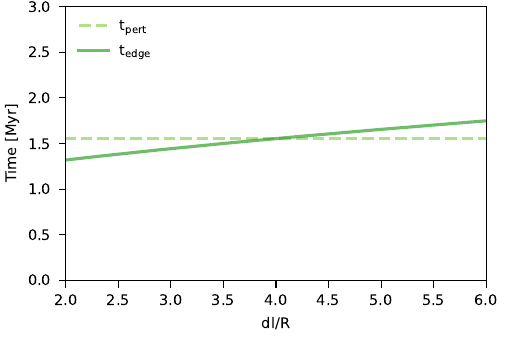}
      \caption{Comparison of the perturbation timescale (Equation \ref{eq:perturbationtimescale}), depicted by the dashed, light green line, with the edge effect formation timescale (Equation \ref{eq:edge} and \ref{eq:accelerationCollapseFilament}), shown by the darker solid, green line.}
      \label{fig:tedge_tpert_example}
  \end{figure}

  To get a criterion for which perturbations would grow faster than the edge effect, we consider a simple geometry: a linear transition from the filament to the outside medium as depicted by the orange line in Figure \ref{fig:gradA} in the first row. We want to determine the density gradient $\rho/dl$ at which perturbations grow as fast as the edge effect $t_{\mathrm{edge}}=t_{\mathrm{pert}}$ as indicated in Figure \ref{fig:tedge_tpert_example}. The dashed line shows the perturbation timescale (Equation \ref{eq:perturbations}) and the solid line is the edge effect formation, given by Equation \ref{eq:edge} using the centre of mass acceleration at the end of a filament for a linear density gradient:
  \begin{align}
      a_{\mathrm{cm}} = -2\uppi G \rho \Biggr[& \sqrt{\frac{\kappa^2}{4}+1} \frac{\kappa R^2}{4dl} - \frac{R^2}{2dl} \tanh ^{-1} \left(\frac{-\kappa/2}{\sqrt{\kappa^2/4+1}}\right) \nonumber \\
      -& \left( \frac{1}{2} + \frac{\kappa R}{4dl} \right) \sqrt{R^2+\left( dl + \frac{\kappa R}{2} \right)^2} + \frac{1}{2} dl \nonumber \\
      +& \frac{R^2}{2dl} \tanh ^{-1} \left( -\frac{dl+\kappa R/2}{\sqrt{(dl+\kappa R/2)^2+R^2}} \right) - l \nonumber \\
      +& \kappa R + \sqrt{R^2 + \left( l - \frac{\kappa R}{2} \right)^2} \Biggr]. \label{eq:acc_cm}
  \end{align}
  The derivation of this equation is provided in Appendix \ref{ap:linear} and is valid for filaments with $l-\kappa R/2 \gg R$, thus filaments with large aspect ratios. The crossing point of the lines indicates the gradient where both effects grow on the same timescale. The determination of this crossing point is not straightforward. Therefore, we use Taylor approximations to determine the critical density gradient. 
  
  In the first step, we do a Taylor expansion of the acceleration at the end of the filament (Equation \ref{eq:acc_cm}) around $dl=0$ with Mathematica: 
  \begin{align}
      a \approx G \uppi \rho \left[ aR+bdl \right],
  \end{align}
  with free $(a,b)$ parameters which are fitted to the numeric values at the end of the section. Inserting this into the edge effect formation timescale leads to:
  \begin{align}
  \label{eq:tedge_approximation}
      t_{\mathrm{edge}} \approx \sqrt{\left( \frac{1}{f}-1\right) \frac{2\kappa R}{G\uppi \rho (aR+bdl)}}. 
  \end{align}
  We did another Taylor expansion around $dl=0$ for the last multiplier:
  \begin{align}
      \frac{1}{\sqrt{(aR+bdl)}} \approx \sqrt{\frac{1}{aR}} - \frac{1}{2} \left[\left( \frac{1}{aR} \right)^{3/2} b \right] dl.
  \end{align}
  Inserting this into the edge effect formation timescale leads to:
  \begin{align}
  \label{eq:tedge_approx}
      t_\mathrm{edge} \approx \sqrt{\left( \frac{1}{f} -1 \right) \frac{2 \kappa R}{G \uppi \rho}} \left[ \sqrt{\frac{1}{aR}} - \frac{1}{2} \frac{b}{(aR)^{3/2}} dl \right].
  \end{align}
  Now the critical density gradient can be determined solving 
  \begin{align}
      t_\mathrm{edge} \overset{!}{=} t_\mathrm{pert}
  \end{align}
  for $dl/R$:
  \begin{align}
      \frac{dl}{R} &\approx - \frac{2a^{3/2}}{b} \left[ t_{\mathrm{pert}} \sqrt{\frac{fG\uppi\rho}{(1-f)2k}} - \frac{1}{\sqrt{a}}\right] \\
      &\approx \alpha \sqrt{\frac{G\uppi \rho f}{(1-f)}} \tau_{\mathrm{dom}} \log \left( \frac{1-f}{f\epsilon} \right) - \beta. \label{eq:gradient}
  \end{align}
  \begin{figure}
      \centering
      \includegraphics[width=\columnwidth]{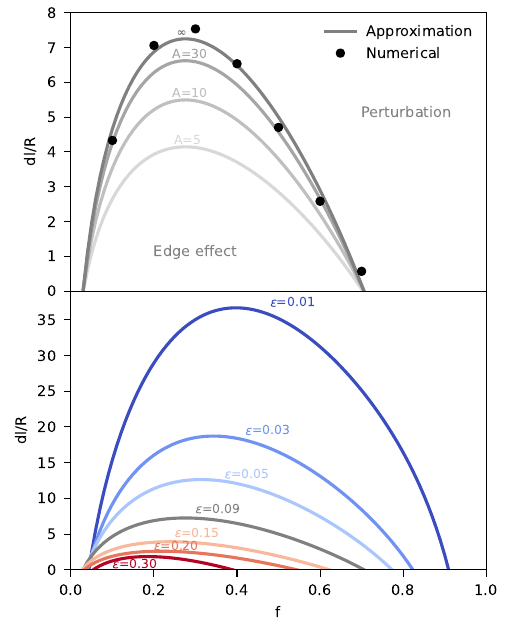}
      \caption{Top panel: The approximation of the critical gradient (Equation \ref{eq:gradient}) is depicted for different criticalities in grey. The different shades depict solutions for different filament aspect ratios, indicated by the given number. The parameters $(\alpha,\beta)$ are fitted to the numerical solutions for a long filament ($R=0.1$\pc, $l=200$\pc), given by the black dots. Below the curve, a dominant edge effect is expected, whereas above the curve, perturbations grow faster, and thus no strong edge effect is expected. \\
      Lower panel: The critical gradient is given for different perturbation strengths $\epsilon$. The grey line shows the observed strength of $\epsilon=0.09$ \citep{Roy2015}, in comparison in blue lower strength and in red stronger ones.}
      \label{fig:gradient_trheshold}
    \end{figure}

  The approximation of the critical gradient is presented in Figure \ref{fig:gradient_trheshold}, upper panel, for different aspect ratios indicated by the shaded grey lines for a perturbation strength of $\epsilon=0.09$ which is the observed value \citep{Roy2015}. The filament's aspect ratio $A$ belonging to the various line is given in the subscript. Our model only holds for long filaments, thus, the solution converges for large values of $A$. Parameters $\alpha=4.22$ and $\beta=8.17$ were fitted to the numeric values for a large aspect ratio ($A=2000$) given as the black dots. The numerical values are well reproduced by the approximation. Filaments below the critical gradient would be expected to show a pronounced edge effect, whereas filaments above the line would grow perturbations on a similar timescale or faster. This shows that for filaments with a criticality greater than 0.7, we would expect perturbations to always grow faster. In the lower panel, we present the critical gradient for different perturbation strengths $\epsilon$. The grey line indicates the observed value of $\epsilon=0.09$ \citep{Roy2015}, which is also used to demonstrate the convergence of the solution. The blue lines present less dominant perturbations whereas the red curves depict the critical gradient for stronger perturbations. As expected for lower perturbations smoother gradients are needed to suppress the edge effect. In addition, the stronger perturbations shift the line-mass $f$ above which all filaments would be dominated by perturbations to lighter filaments and to heavier filaments for the less perturbed filaments.

  \begin{figure}
      \centering
      \includegraphics[width=\columnwidth]{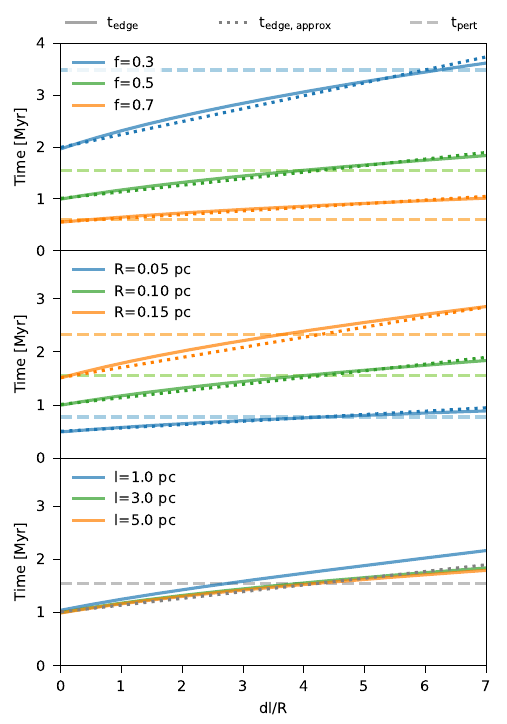}
      \caption{The same as Figure \ref{fig:tedge_tpert_example}, but adding a comparison to the approximated edge effect time (Equation \ref{eq:tedge_approx}) with the dotted lines. In the first row, the dependency on criticality is depicted, in the second row on radius and in the last on the filament length. At the crossing points between $t_{\mathrm{edge}}$ and $t_{\mathrm{pert}}$ the approximation of the edge effect formation timescale is reasonable.}
      \label{fig:tedge_tpert}
  \end{figure}

  For validation, we compare the crossing point of $t_\mathrm{edge}$ and $t_\mathrm{pert}$ (see Figure \ref{fig:tedge_tpert_example}) to the approximation of the edge effect formation timescale given by Equation \ref{eq:tedge_approx} in  Figure \ref{fig:tedge_tpert}. As in Figure \ref{fig:tedge_tpert_example}, the dashed line denotes $t_\mathrm{pert}$, the solid line $t_\mathrm{edge}$ for a filament with linear density gradient and the dotted line its approximation $t_\mathrm{edge,approx}$, given by Equation \ref{eq:tedge_approximation}. The criticality is varied in the top row, the radius in the middle and the length in the last one, respectively indicated by different colours. As we used the approximation to determine the critical gradient the approximation has to be valid for the intersections, which fits well for the presented cases.
  
  \begin{figure}
      \centering
      \includegraphics{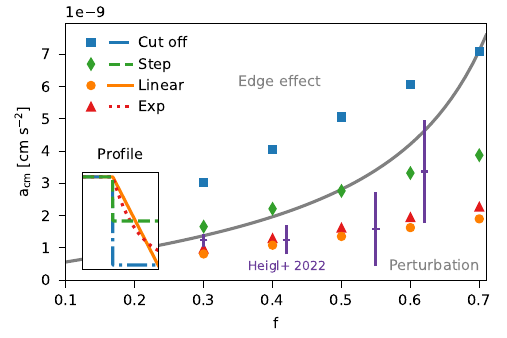}
      \caption{The grey line depicts the critical centre of mass acceleration (Equation \ref{eq:acc_crit}) depending on criticality. The different markers give numeric results of the centre of mass acceleration of different filament end profiles given in the subplot. Above the grey line the edge effect is expected to be dominant, whereas below, perturbations would grow faster. In comparison, the violet crosses show the acceleration measured in the simulation by \citet{Heigl2022}, where the edge effect was suppressed in agreement with our model. The used parameters were: $\epsilon=0.09$, $R=0.02$\pc, $l=0.94$\pc, $dl/R=9$.}
      \label{fig:critical_acceleration}
  \end{figure}
  
  Altogether, the density gradient at the end of the filament leads to a decrease in acceleration, thus, there also has to be a critical acceleration which is needed in a certain setup to have dominant end cores. This can be calculated without any additional approximations using the same ansatz as before and solving for the acceleration:
  \begin{align}
      t_\mathrm{edge} \overset{!}{=} t_\mathrm{pert}
  \end{align}
  \begin{align}
      a_{\mathrm{crit}} = \left( \frac{1}{f} -1 \right)\frac{2\kappa R}{\tau _{\mathrm{dom}}} \log ^{-2} \left( \left( \frac{1}{f} - 1 \right) \frac{1}{\epsilon}\right) .\label{eq:acc_crit}
  \end{align}
  
  An example is given in Figure \ref{fig:critical_acceleration} as a grey line for $\epsilon=0.09$, $R=0.02$\pc, $l=0.94$\pc, $dl/R=9$. The markers indicate the numerically determined acceleration of the centre of mass of the filament's end region for different profiles, indicated in the subplot. The violet crosses show the values of a simulation by \citet{Heigl2022} where the filament is created by a converging flow. While the filament starts to collapse material comes in at the filament's ends due to the constant inflow region. This creates a linear density gradient in the end region. Thus, it fits very well with our numeric solution for a linear end. Above the critical acceleration (grey line) an edge effect is expected whereas below perturbations should grow faster. Since the acceleration varies strongly in the end region, the determination from the simulation is error-prone, thus we show an error bar of the standard deviation 10 voxels ahead of the peak and behind. The linear and exponential profiles, which have a similar gradient, lie below the line for this specific example, whereas the hard cut-offs are expected to have dominant ends, especially for low line-masses. Since \cite{Heigl2022} did also not detect dominant end cores, it fits our results.
  
\section{Validation by simulations}
\label{sec:simulation}
 
  For validation, we performed isothermal hydrodynamic simulations with the adaptive mesh refinement code RAMSES \citep{Teyssier_2002}. With a second-order Gudonov solver the Euler equations are solved in their conservative form using a  MUSCL \citep[Monotonic Upstream-Centered Scheme for Conservation Laws,][]{Leer1979}, the HLLC-Solver \citep[Harten-Lax-van Leer- Contact,][]{Toro1994} and the MC slope limiter \citep[monotonized central-differenc,][]{Leer1979}.
 
\begin{figure}
     \centering
     \includegraphics{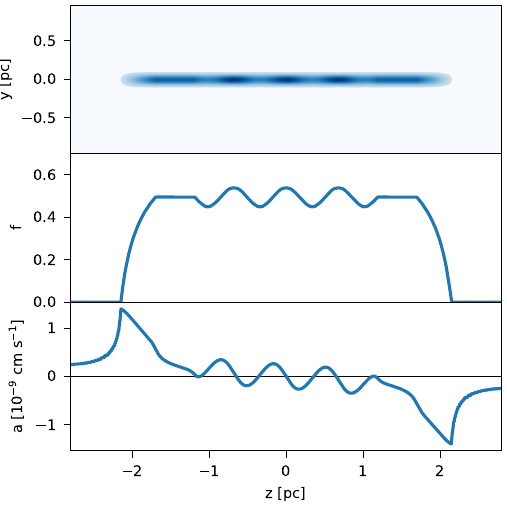}
     \caption{An exemplary simulation environment to validate the critical gradient, in this case with $R=0.1$\pc, $f=0.5$ and $dl/R=6$. The upper part shows a section of a density slice, the second plot the criticality along the z-axes and the last one the gravitational acceleration along the z-axes. Inside the filament perturbations with $\epsilon=0.09$ are set up followed by a flat profile in the end region and a linear density gradient, which in this example is $dl/R=6$. Since there is a density cut to the ambient medium in the simulations at the filament boundary density $\rho_\mathrm{b}$, the gradient is curved in the line-mass projection.}
     \label{fig:simulation_setup}
  \end{figure}

  We set up filaments with perturbations in the line-mass as displayed in Figure \ref{fig:simulation_setup}. In the upper panel, a density slice through the filament is given and in the lower panels the according criticality and acceleration along the z-axes. The bending of the gradient in the end region is due to the cut off at boundary density, which will be discussed at the end of this section. Perturbations were inserted via Equation \ref{eq:perturbations} going over into a flat region at the end, to not disturb the edge effect. At the filament end the linear transition into the surrounding density is depicted, in this case with a gradient of $dl/R=6$. 
  We cut the outermost peaks to not have an overlap between perturbations and the edge effect. Then, we varied the linear density gradient at both ends to validate the transition between perturbations and edge effect as given by Equation \ref{eq:gradient}. We consider perturbations to be dominant if all cores collapse within a time span of 10\% of each other, or the inner cores collapse first since then no pronounced edges can be detected. In contrast, the edge effect is considered to be dominating if the end cores collapse before the inner cores.  
 
\begin{figure}
      \centering
      \includegraphics{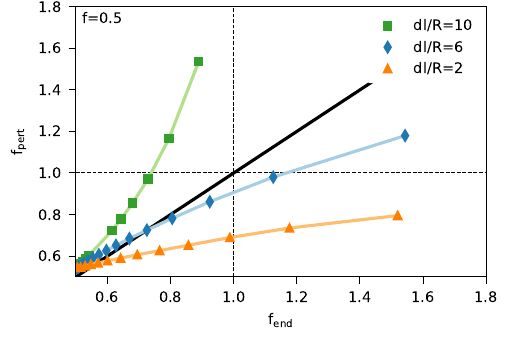}
      \caption{For simulations with $f=0.5$, $R=0.1$\pc\ the evolution of the criticality for the end cores against the criticality of the perturbations are shown for 3 different gradients. All simulations start at $f_{\mathrm{pert}}=f_{\mathrm{edge}}=0.5$ and move to larger maximal f-values with time. Each marker represents one simulation time-step ($\Delta t=0.1$\Myr). Above the solid black line, perturbations grow faster in f than the edge. A clear edge effect can only be seen for $dl/R=2$, for the others perturbations grow on similar timescales or faster, which fits the prediction.}
      \label{fig:f_evolution}
  \end{figure}

  As a first test, we performed simulations for $f=0.5$ and different gradients, given in Figure \ref{fig:f_evolution}. Here we plot the criticality of the cores at the end $f_\mathrm{end}$ against the criticality of the cores formed from perturbations $f_\mathrm{pert}$. Every marker is one output of the simulation in time-steps of $\Delta t=0.1$\Myr\ with $R=0.1$\pc, $l=0.3$\pc\ and $f=0.5$. For $dl/R=10$ and $dl/R=6$, both cores collapse within one time-step, thus, perturbations and edge effect collapse on the same timescale and no clear edge effect can be detected. Only for $dl/R=2$ the edge effect collapses clearly faster than the perturbations which agrees with the fact that in this case a dominant edge core would be expected. This validates the critical gradient given in Figure \ref{fig:gradient_trheshold} which predicts the transition between end core and perturbation domination for these parameters to be at $dl/R=5$. Further validations for different parameters can be found in Table \ref{tab:validation}.
 
 \begin{table} 
      \centering
      \caption{Results of the simulation "sim" in comparison to the expected result by the model "exp". "e" means simulations with dominant edge effect or "p" configurations where perturbations grow on the same timescale or faster.  The edge effect and perturbation timescale, as determined in the simulation, are given in comparison to their expected values in brackets. An output timestep in the simulation takes $\Delta t = 0.1$\Myr. The length was always given by five times the dominant wavelength (Equation \ref{eq:lambda}).}
      \label{tab:validation}
      \begin{tabular}{cccccccc}
        f   &   R [pc] &   dl/R   & $t_{\mathrm{edge}}$ [Myr] & $t_{\mathrm{pert}}$ [Myr] &  exp &   sim \\
        \hline
        0.5 & 0.1 & 10 & 1.76 (2.02) & 1.57 (1.55) & p & p \\ 
        0.5 & 0.1 & 6 & 1.47 (1.71) & 1.57 (1.55) & p & p \\
        0.5 & 0.1 & 2 & 1.18 (1.31) & 1.57 (1.55) & e & e \\
        0.5 & 0.05 & 6 & 0.78 (0.84) & 0.78 (0.78) & p & p \\
        0.3 & 0.05 & 10 & 1.27 (1.94) & 1.67 (1.74) & p & e \\
        0.3 & 0.1 & 10 & 2.55 (3.99) & 3.14 (3.49) & p & e \\
      \end{tabular}
  \end{table}

\begin{figure}
      \centering
      \includegraphics{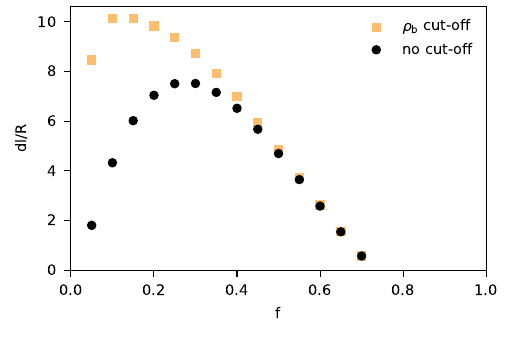}
      \caption{Critical gradient for a cut off at $\rho_\mathrm{b}$ given by the squares in light orange and the former solution without any cut off in comparison as black dots. For line-masses $f \geq 0.4$ the deviations are small. However, for lighter filaments, the estimation is not sufficient an much smoother gradients are necessary.}
      \label{fig:critical_gradient_corr}
  \end{figure}

  As a second test, we did simulations with smaller line-masses, given in Table \ref{tab:validation}. For low line-masses the predicted timescales for the edge effect deviate from the simulated ones. For large line-mass filaments this has only a minor impact on the prevalent fragmentation mode, however, for lighter filaments, the deviations are so strong that the edge effect is still dominant for density gradients above the critical one. This is due to the fact that we consider the density gradient to extend to zero in the analytic approach. However, we expect observed filaments to be constrained by an outside pressure. This is also the case in the simulations meaning that the gradient is cut off at a boundary density $\rho_\mathrm{b}$ which is in pressure equilibrium with the surrounding material (see Figure \ref{fig:appendix} dashed-dotted line). This leads to a loss in mass in the end region which increases the acceleration at the end and results in shorter collapse timescales than predicted. Since the acceleration can be considered constant during the edge formation, a deviation in acceleration has a stronger impact on longer timescales. Figure \ref{fig:critical_gradient_corr} shows an approximation of how a cut-off at the boundary density would influence the critical density gradient, given by the orange squares (the new acceleration used to determine $t_{\mathrm{edge}}$ is derived in Appendix \ref{ap:linear} and given by Equation \ref{eq:a_cm'}). This is only an approximation since a cut-off at the boundary density not only constrains the filament along its main axis but also reduces the radius gradually in the end region, as can be seen in Figure \ref{fig:simulation_setup}. This was not accounted for in the comparison of Figure \ref{fig:critical_gradient_corr}. Thus, for low line-mass filaments, even shallower density gradients than predicted are necessary to suppress the edge effect. However, the linear gradient is a good first approximation, especially for filaments with $f\geq 0.4$.

\section{Discussion}
\label{sec:discussion}
  
  For the calculation of the critical gradient, several approximations have been made. However, Figure \ref{fig:tedge_tpert} already showed a comparison between the numerical collapse times and the approximated collapse times. For the tested cases, which show a reasonable span over the parameter space of filaments the crossing point of the two timescales is approximated well. In addition, the determined curve of the critical gradient fits well with the numerically calculated curve (Figure \ref{fig:gradient_trheshold}). Although we observe a systematic deviation for filaments with lower aspect ratio $A$, which is expected due to the used approximation, the values do converge toward larger aspect ratios. However, the deviation is limited and can be interpreted as a transition region between the two effects. Thus, the approximations seem applicable to our case.
  
  Whether the size of the end region $dl$ is realistic is difficult to say. Although there are some observations of the line-mass distribution along filaments as \citet{Roy2015,Kainulainen2016,Cox2016,Yuan2020,Schmiedeke2021}, the end region is often not displayed or not enough resolved to make a reasonable statement about the gradient. The best example is the observation of Barnard 5 by \citet{Schmiedeke2021}. Here a clear linear trend is seen at the end of the filament. The declining region is of the order of $\sim0.05$\pc, whereas the FWHM is $\sim0.03\pc$, thus, in this case, the end region is indeed bigger than the filament radius, which is needed for perturbations to be as fast as the edge effect in the parameter space we investigated. However, the filament is highly supercritical and thus, outside the range where our model is applicable. Further observations of filament end regions are necessary to make a reasonable comparison and to test our prediction.
  
\section{Conclusion}
\label{sec:conclusion}
    
  We showed that a density gradient at the end of a filament cannot stop the edge effect, since every kind of density gradient will cause a gradient in acceleration. The gradient will, however, only cause a weakening of the acceleration which leads to a longer edge effect formation timescale. But since the collapse timescale of a filament scales with the same respect to $a$ the edge effect will take longer but will nevertheless occur before the collapse.
  
  However, perturbations are independent of the acceleration at the end of the filament, and thus if the edge effect is slowed down significantly, perturbations can grow on similar timescales or even faster. Assuming the density gradient at the end of the filament to be linear, we presented the critical density gradient which is needed for perturbations to be dominant in the filament. For filaments beyond $f=0.7$, perturbations always grow faster for the observed perturbation strength of 9\% or larger. For lower line-mass filaments, an end region of several times the radius in length is needed to slow down the edge effect significantly, such that perturbations grow faster. However, deviations from the model are expected for $f<0.4$ due to the approximations in the derivation of the critical gradient, where even shallower gradients would be needed. Altogether, density gradients at the end of the filament could be the reason, why the edge effect is only rarely observed. Observations are however needed to show whether such density gradients in end regions exist.

\section*{Acknowledgements}

   This research was supported by the Excellence Cluster ORIGINS which is funded by the Deutsche Forschungsgemeinschaft (DFG, German Research Foundation) under Germany’s Excellence Strategy - EXC-2094 - 390783311. We thank the CAST group for the helpful discussion and comments, as well as the anonymous referee for the valuable feedback.

\section*{Data Availability}

The data will be made available on request.



\bibliographystyle{mnras}
\bibliography{Literature} 




\appendix

\section{Derivation of the accelerations for different profiles and their gradients} \label{ap:derivation_gradients}

  Consider a uniform density distribution in the radial direction of the filament. The acceleration is determined by integrating over the density distribution:
  \begin{align}
      a &= G \int_0 ^R \mathrm{d}r \int_0^{2\uppi}\mathrm{d}\theta \int_{\mathrm{Fil.}}\mathrm{d}z' \rho (z') \frac{r}{\sqrt{r^2+z'^2}^2} \cos(\alpha) \\
      &= 2\uppi G \int_0 ^R \mathrm{d}r \int_{\mathrm{Fil.}}\mathrm{d}z' \rho (z') \frac{rz'}{\left(r^2+z'^2\right)^{3/2}}
  \end{align}
  Substitution $u=r^2$ $\frac{du}{dr}=2r$
  \begin{align}
      a &= \uppi G \int_0 ^{R^2} \mathrm{d}u  \int_{\mathrm{Fil.}}\mathrm{d}z' \rho (z') \frac{z'}{\left(u+z'^2\right)^{3/2}} \\
      &= -2\uppi G \int_{\mathrm{Fil.}}\mathrm{d}z' \rho (z') \left[ \frac{z'}{\sqrt{ R^2+z'^2 }} - \frac{z'}{\sqrt{z'^2}} \right] \label{eq:acc_int}.
  \end{align}
  Since this function is not continuous at $z=0$ the limits have to be adjusted.
  
  \subsection{Hard cut}\label{ap:hard_cut}

  \begin{figure}
      \centering
      \includegraphics{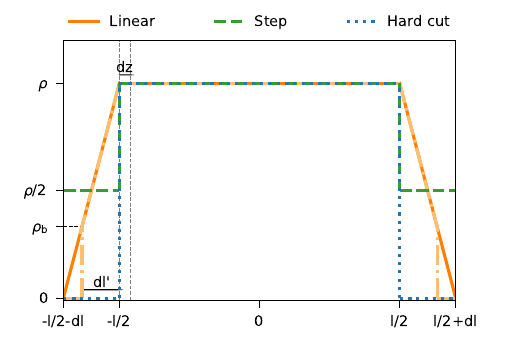}
      \caption{Schematic drawing of the density distribution for the three different configurations (linear, step and hard cut) of the end regions. In addition, the linear case with cut-off at the boundary density is indicated by the light orange dashed-dotted line.}
      \label{fig:appendix}
  \end{figure}

  For a hard cut-off, the density distribution is simple, see Figure \ref{fig:appendix} blue dotted line. Integrating Equation \ref{eq:acc_int} leads to the acceleration from \cite{Burkert2004}:
  \begin{align}
      a =& - 2 \uppi G \rho \Bigg[ \int_{-l/2-z}^{l/2-z} \mathrm{d}z' \frac{z'}{\left( R^2+z'^2 \right)^{1/2}} \nonumber \\
      &- \int_{-l/2-z}^{0} \mathrm{d}z' (-1) - \int_0^{l/2-z} \mathrm{d}z' (+1) \Bigg] \\
      =& -2\uppi G \rho \left[ \sqrt{R^2+(l/2-z)^2} - \sqrt{R^2+(l/2+z)^2} + 2z \right].
  \end{align}
  Simple differentiation leads to the acceleration gradient:
  \begin{align}
      \nabla a =& -2\uppi G \rho \frac{\mathrm{d}}{\mathrm{d}z} \left[ \sqrt{R^2+(l/2-z)^2} - \sqrt{R^2+(l/2+z)^2} + 2z \right]\\
      =&-2\uppi G \rho \left[ -\frac{l/2+z}{\sqrt{R^2+(l/2+z)^2}} - \frac{l/2-z}{\sqrt{R^2+(l/2-z)^2}} + 2 \right].
  \end{align}
  
  \subsection{Step}\label{ap:step}

  For the step function integration over the density distribution given in Figure \ref{fig:appendix} by the green dashed line leads to the acceleration along such a filament
  \begin{align}
      a =& -2 \uppi G \Bigg[ \int_{-l/2-dl-z}^{-l/2-z} \mathrm{d}z' \frac{\rho}{2} \left( \frac{z'}{\left( R^2 + z'^2 \right)^{1/2}} -1 \right) \\ &+ \int_{-l/2-z}^0 \mathrm{d}z' \rho \left( \frac{z'}{\left(R^2+z'^2 \right)^{1/2}} -1 \right) \nonumber \\
      &+ \int_{0}^{l/2-z} \mathrm{d}z' \rho \left( \frac{z'}{\left(R^2+z'^2 \right)^{1/2}} +1 \right) \nonumber \\
      &+\int_{l/2}^{l/2+dl/2-z} \mathrm{d}z' \frac{\rho}{2} \left( \frac{z'}{\left( R^2 + z'^2 \right)^{1/2}} +1 \right) \Bigg] \\
      &= -2\uppi G \rho \Bigg[ -\frac{1}{2} \sqrt{R^2+(l/2+z)^2} + \frac{1}{2} \sqrt{R^2+(l/2-z)^2} \nonumber \\
      &+ 2z -\frac{1}{2} \sqrt{R^2+(l/2+dl+z)^2} + \frac{1}{2} \sqrt{R^2+(l/2+dl-z)^2} \Bigg].
  \end{align}
  From this the acceleration gradient follows as
  \begin{align}
      \nabla a =& -2 \uppi G \rho \Bigg[ -\frac{1}{2} \frac{l/2+z}{\sqrt{R^2+(l/2+z)^2}} - \frac{1}{2} \frac{l/2-z}{\sqrt{R^2+(l/2-z)^2}} \nonumber \\
      & + 2- \frac{1}{2} \frac{l/2+dl+z}{\sqrt{R^2+(l/2+dl+z)^2}} - \frac{1}{2} \frac{l/2+dl-z}{\sqrt{R^2+(l/2+dl-z)^2}}\Bigg].
  \end{align}

  \subsection{Linear}\label{ap:linear}

  Calculating the acceleration gradient for a filament with a linear end region is a bit more complex than in the previous cases. Thus, we only calculated it for the exact end of the filament at $z=l/2$ building the difference quotient at location $-l/2$ and $-l/2+dz$ as indicated by the grey dashed lines in Figure \ref{fig:appendix}. The contribution of the linear part is given by integrating with a liner density distribution $\rho (z') = mz'+b$:
  \begin{align}
      I(z') =& \int \mathrm{d}z'  (mz'+b) \left( \frac{z'}{\sqrt{R^2+z'^2}} + 1\right) \\
      =& \Bigg[ \frac{m}{2} \left( z' \sqrt{R^2+z'^2} - R^2 \tanh^{-1}\Bigg({\frac{z'}{\sqrt{R^2+z'^2}}}\Bigg) \right) \nonumber \\
      &+ \frac{m}{2} z'^2 + b \sqrt{R^2+z'^2} + bz' \Bigg] \\
      =& \left[ \left( \frac{m}{2} z' + b \right) \left( \sqrt{R^2+z'^2} + z' \right) - \frac{R^2m}{2} \tanh^{-1}\Bigg({\frac{z'}{\sqrt{R^2+z'^2}}}\Bigg) \right].
  \end{align}
  As before the contribution of the inner part is given by the acceleration of the constant-density filament:
  \begin{align}
      a(z) = \rho \left[ \sqrt{R^2+\left(\frac{l}{2}-z\right)^2} - \sqrt{R^2+\left(\frac{l}{2}+z\right)^2} + 2z \right] .
  \end{align}
  To determine the gradient of the acceleration we use these density distributions at the filaments end and $dz$ infinitesimal shifted from the end:
  \begin{align}
      \rho_1 (z) &= \rho \left( \frac{1}{dl} z + 1 \right), \\
      \rho_2 (z) &= \rho \left( \frac{1}{dl} z + 1 + \frac{dz}{dl} \right).
  \end{align}
  Since we are only calculating the gradient for the end of the filament, the contributions from the other end can be neglected in this case, which leads to the following difference quotient:
  \begin{align}
      \Delta a =& \frac{2\uppi G}{dz} \left[ I|_{-dl}^{0} + a(-l/2) - I|_{-dl-dz}^{-dz} -a\left(-\frac{l}{2}+dz \right) \right] \\
      =& \frac{2\uppi G \rho}{dz} \bigg[ R - \frac{1}{2} \left( \sqrt{R^2+dl^2} -dl \right) + \frac{R^2}{2dl} \tanh^{-1}\Bigg({\frac{-dl}{\sqrt{R^2+dl^2}}}\Bigg) \nonumber \\
      &+ \sqrt{R^2+l^2} - R -l - \left( \frac{-dz}{2dl} + 1 + \frac{dz}{dl} \right) \left( \sqrt{R^2+dz^2} - dz \right) \nonumber \\
      &+ \frac{R^2}{2dl} \tanh^{-1}\Bigg({\frac{-dz}{\sqrt{R^2+dz^2}}}\Bigg) + \left( \frac{1}{2} + \frac{dz}{2dl} \right) \biggl( \sqrt{R^2+(dl+dz)^2} \nonumber \\
      &- dl - dz \biggl) - \frac{R^2}{2dl} \tanh^{-1}\Bigg({\frac{-dl-dz}{\sqrt{R^2+(dl+dz)^2}}}\Bigg) - \sqrt{R^2+(l-dz)^2} \nonumber \\
      &+ \sqrt{R^2+dz^2} + l - 2dz  \bigg] \\
      =&  \frac{2 \uppi G \rho}{dz} \bigg[ -\frac{1}{2} \sqrt{R^2+dl^2} + \frac{dl}{2} + \frac{R^2}{2dl} \tanh^{-1}\Bigg({\frac{-dl}{\sqrt{R^2+dl^2}}}\Bigg) \nonumber \\
      &+ \sqrt{R^2+l^2} - \frac{dz}{2dl} \sqrt{R^2+dz^2} + \frac{dz^2}{2dl} - \sqrt{R^2+dz^2} + dz \nonumber \\
      &+ \frac{R^2}{2dl} \tanh^{-1}\Bigg({\frac{-dz}{\sqrt{R^2+dz^2}}}\Bigg) + \frac{dz}{2dl} \sqrt{R^2+(dl+dz)^2} - \frac{dz}{2} \nonumber \\
      &- \frac{dz^2}{2dl} + \frac{1}{2} \sqrt{R^2+(dl+dz)^2} - \frac{dl}{2} - \frac{dz}{2}  \nonumber \\
      &- \frac{R^2}{2dl} \tanh^{-1}\Bigg({\frac{-dl-dz}{\sqrt{R^2+(dl+dz)^2}}}\Bigg) - \sqrt{R^2+(l-dz)^2} \nonumber \\
      &+ \sqrt{R^2+dz^2} - 2dz \bigg].
  \end{align}
  The gradient is then given for $dz\to0$:
  \begin{align}
      \nabla a =& \lim_{dz\to0} \Delta a \\
      =& 2 \uppi G \rho \bigg[ \frac{\sqrt{R^2+dl^2}}{dl} - \frac{R}{dl} + \frac{l}{\sqrt{R^2+l^2}} - 2 \bigg].
  \end{align}
  For determining the edge effect formation timescale for a linear end region (e.g. Figure \ref{fig:tedge_tpert}, and the critical density gradient), the centre of mass acceleration is needed. The centre of mass of the end region is again considered as $-l/2+\kappa R/2$, so the density distribution is given by:
  \begin{align}
      \rho_{\mathrm{cm}} (z) = \rho \left(\frac{1}{dl} + 1 + \frac{kR}{2dl} \right)
  \end{align}
  with this, the center of mass acceleration follows
  \begin{align}
      a_{\mathrm{cm}} =& - 2\uppi G \left[ I|_{-\kappa R/2-dl}^{-\kappa R/2} + a\left(-\frac{l}{2}+ \frac{\kappa R}{2} \right) \right] \\
      =& -2 \uppi G \rho_0 \bigg[ \left( \frac{\kappa R}{4dl} + 1 \right) \left( \sqrt{R^2 + (\kappa R/2)^2} - \frac{\kappa R}{2} \right) \nonumber \\
      &- \frac{R^2}{2dl} \tanh^{-1}\Bigg({\frac{-\kappa R/2}{\sqrt{R^2+ (\kappa R /2)^2}}}\Bigg) + \sqrt{R^2 + (l - \kappa R /2)^2} \nonumber \\
      &- \sqrt{R^2 + (\kappa R/2)^2} + 2 \left( -\frac{l}{2} + \frac{\kappa R}{2} \right) - \left( \frac{\kappa R}{4dl} + \frac{1}{2} \right) \nonumber \\
      &\left( \sqrt{R^2+ (\kappa R/2+dl)^2} -\frac{\kappa R}{2} - dl \right) \nonumber \\
      &+ \frac{R^2}{2dl} \tanh^{-1}\Bigg({\frac{-\kappa R/2 - dl}{\sqrt{R^2+ (\kappa R/2 + dl)^2}}}\Bigg) \Bigg] \\
      =& -2 \uppi G \rho_0 \Bigg[ \frac{\kappa R}{4dl} \sqrt{R^2 + (\kappa R/2)^2} \nonumber \\
      &- \frac{R^2}{2dl} \tanh^{-1}\Bigg({\frac{-\kappa R/2}{\sqrt{R^2+(\kappa R/2)^2}}}\Bigg) + \sqrt{R^2+(l-\kappa R/2)^2} -l \nonumber \\
      &+\kappa R - \left( \frac{1}{2} + \frac{\kappa R}{4dl} \right) \sqrt{R^2+(\kappa R/2 + dl)^2} \nonumber \\
      &+ \frac{dl}{2} + \frac{R^2}{2dl} \tanh^{-1}\Bigg({\frac{-\kappa R/2 - dl}{\sqrt{R^2+ (\kappa R/2+dl)^2}}}\Bigg) \Bigg] . \label{eq:a_cm}
  \end{align}
  Approximation for $l-\kappa R /2 \gg R$ leads to
  \begin{align}
      a_{\mathrm{cm}} =& -2\uppi G \rho \Bigg[ \frac{\kappa R}{4dl} \sqrt{R^2+(\kappa R/2)^2} - \frac{R^2}{2dl} \tanh^{-1}\Bigg({\frac{-\kappa R/2}{\sqrt{R^2+(\kappa R/2)^2}}}\Bigg) \nonumber \\
      &+ \frac{\kappa R}{2} - \Bigg( \frac{1}{2} - \frac{\kappa R}{4dl} \Bigg) \sqrt{R^2 + (\kappa R/2+dl)^2} +\frac{dl}{2}  \nonumber \\
      &+ \frac{R^2}{2dl} \tanh^{-1}\Bigg({\frac{-\kappa R/2 -dl}{\sqrt{R^2 + (\kappa R/2 +dl)^2}}} \Bigg) \Bigg] .
  \end{align}
  In order to compare the analytic model to simulations and observations, it has to be considered that filaments are surrounded by an outside pressure which cuts off the density profile at the pressure equilibrium at boundary density $\rho_\mathrm{b}$. Thus, the linear profile only extends to $dl'$ while still having a gradient of $m=\rho /dl$ (see light orange dashed-dotted line in Figure \ref{fig:appendix}). The centre of mass acceleration is then determined by:
  \begin{align}
      a_{\mathrm{cm}} =& - 2\uppi G \left[ I|_{-\kappa R/2-dl'}^{-\kappa R/2} + a\left(-\frac{l}{2}+ \frac{\kappa R}{2} \right) \right] \\
      =& 2\upi G \rho \bigg[ \left( \frac{\kappa R}{4dl} + 1 \right) \left( \sqrt{R^2 + \left( \frac{\kappa R}{2}\right)^2} -\frac{\kappa R}{2}\right) \nonumber \\
      &- \frac{R^2}{2dl} \tanh^{-1}\Bigg({\frac{\kappa R / 2}{\sqrt{R^2 + (\kappa R/2)^2}}}\Bigg) -\left( \frac{\kappa R}{4 dl} - \frac{dl'}{2dl} + 1 \right) \nonumber \\
      &\left( \sqrt{R^2 + \left( \frac{\kappa R}{2} + dl' \right)^2} - \frac{\kappa R}{2} - dl'\right) \nonumber \\
      &+ \frac{R^2}{2dl} \tanh^{-1}\Bigg({\frac{-\kappa R /2 - dl'}{\sqrt{R^2 + (\kappa R/2 + dl')^2}}}\Bigg) \nonumber \\
      &+ \sqrt{R^2 + \left( l - \frac{\kappa R}{2} \right)^2} - \sqrt{R^2 + \left( \frac{\kappa R}{2} \right)^2} - l + \kappa R \bigg] \\
      =& 2 \upi G \rho \bigg[ \frac{\kappa R}{4 dl} \sqrt{R^2 + \left( \frac{\kappa R}{2}\right)^2} - \frac{R^2}{2dl} \tanh^{-1}\Bigg({\frac{\kappa R/2}{\sqrt{R^2+(\kappa R/2)^2}}}\Bigg) \nonumber \\
      &- \left( \frac{\kappa R}{4dl} - \frac{dl'}{2dl} + 1 \right) \sqrt{R^2 + \left( \frac{\kappa R}{2} + dl' \right)^2} - \frac{dl'^2}{2dl} + dl' \nonumber \\
      &+ \frac{R^2}{2dl} \tanh^{-1}\Bigg({\frac{-\kappa R/2-dl'}{\sqrt{R^2+ (\kappa R/2+ dl')^2}}}\Bigg) \nonumber \\
      &+ \sqrt{R^2 + \left( l - \frac{\kappa R}{2} \right)^2} - l + \kappa R \bigg] .\label{eq:a_cm'}
  \end{align}
  For $dl'=dl$ this turns into Equation \ref{eq:a_cm} since then $\rho _\mathrm{b} = 0$.


\bsp	
\label{lastpage}
\end{document}